\documentclass[11pt]{article}
\usepackage{putex}

\begin{document}
\preprint{PUPT-2464}

\title{An axial gauge ansatz for higher spin theories}
\institution{PU}{Joseph Henry Laboratories, Princeton University, Princeton, NJ 08544, USA}
\authors{Steven S. Gubser and Wei Song}
\abstract{
We present an ansatz which makes the equations of motion more tractable for the simplest of Vasiliev's four-dimensional higher spin theories.  The ansatz is similar to axial gauge in electromagnetism.  We present a broad class of solutions in the gauge where the spatial connection vanishes, and we discuss the lift of one of these solutions to a full spacetime solution via a gauge transformation.
}
\date{November 2014}

\maketitle

\section{Introduction}
\label{INTRODUCTION}

Vasiliev's higher spin theories in four dimensions \cite{Vasiliev:1990en,Vasiliev:1992av} are relatively simple theories involving infinitely many fields, all with integer spin.  The full non-linear equations of motion are known, and the simplest solution to them is $AdS_4$.  Some additional solutions of the equations \eno{eoms} are known: See for example \cite{Sezgin:2005pv,Didenko:2009td,Iazeolla:2012nf}.  Finding exact solutions is challenging because the equations of motion are non-linear and involve a non-local star product in the oscillator variables.  But a broader set of exact solutions is highly desirable in order to advance our understanding of classical higher spin theory beyond perturbation theory.  The aim of this paper is to introduce a new class of exact solutions.  In one subcase of our construction, the solutions are parametrized by an arbitrary function of three variables, making it a remarkably large class of solutions.

Vasiliev's equations involve auxiliary, bosonic, spinorial variables $z^\alpha$, and one of the equations of motion takes the form $f_{z^1z^2} = -p(b * K)$, where $f_{z^1z^2}$ is like a Yang-Mills field strength, $b * K$ is covariantly constant in the adjoint representation, and $p$ is a phase---for our purposes, either $1$ or $i$.  The equation $f_{z^1z^2} = -p(b * K)$ is formally similar to having a magnetic field in two dimensions: $\partial_1 A_2 - \partial_2 A_1 = B_{12}$.  A standard strategy is to set $A_1 = 0$ as a gauge choice and then solve for $A_2$ in terms of $B_{12}$.  This is axial gauge.  We are going to make an analogous ansatz, namely $s_1=0=\bar{s}_{\dot{1}}$ where $s_\alpha$ is the spinorial part of the gauge potential with field strength $f_{z^1z^2}$, and $\bar{s}_{\dot\alpha}$ corresponds to a conjugate field strength $f_{\bar{z}^{\dot{1}}\bar{z}^{\dot{2}}}$.  This choice appears to be as innocuous as the choice of axial gauge; however, our overall ansatz is more restrictive than just a gauge choice.

Setting $s_1 = \bar{s}_{\dot{1}} = 0$ removes some star-(anti)-commutators from the equations of motion, so that some components of these equations become linear.  After solving these linear equations (in a gauge where the spacetime components of the higher spin connection vanish), we find that the non-linear equations reduce to quadratic constraints on the ansatz.  These quadratic constraints have many solutions, especially in a particular case where a principle of superposition operates, allowing us to construct solutions labeled by the aforementioned arbitrary function of three variables.  Related strategies have been pursued in previous work \cite{Iazeolla:2007wt,Didenko:2009td}; a common thread is rendering the equation for $f_{z^1z^2}$ effectively linear.

The structure of the rest of the paper is as follows.  For the sake of a self-contained presentation, we review in section~\ref{EOMS} the equations of motion of the higher-spin theories that we are going to solve.  In section~\ref{ANSATZ} we explain in detail the ansatz and show some examples of solutions.  The treatment of this section relies entirely on a gauge where the spacetime components of the connection vanish, also described as the $Z$-space approach in \cite{Sezgin:2005pv}.  In section~\ref{GAUGE} we discuss how solutions of the type obtained in the previous section can be lifted via a gauge transformation to full spacetime solutions.  We focus on a particular route to the Poincar\'e patch of $AdS_4$, but a different gauge transformation would lead to global $AdS_4$.  An example presented in section~\ref{EXAMPLE} leads to an exact solution of the Vasiliev equations in which the spatial part of the higher spin connection is the same as in $AdS_4$ and the scalar takes a form which, in the linearized theory, is associated with a massive deformation of the $O(N)$ model.  It is tempting to identify the exact solution as dual to the massive $O(N)$ model; however, we caution that the explicit breaking of Lorentz symmetry inherent in our ansatz complicates this interpretation.

\section{The equations of motion}
\label{EOMS}

The equations of motion of Vasiliev's higher spin theories in four dimensions \cite{Vasiliev:1990en,Vasiliev:1992av} can be stated in terms of a gauge field
 \eqn{Fields}{
  A = W_\mu dx^\mu + S_\alpha dz^\alpha + \bar{S}_{\dot\alpha} d\bar{z}^{\dot\alpha}
 }
and a scalar field $B$: following the conventions of \cite{Giombi:2010vg}, one writes
 \eqn{eoms}{
  F &\equiv dA + A * A = p (B * K) dz^2 + \bar{p} (B * \bar{K}) d\bar{z}^2  \cr
  DB &\equiv dB + A * B - B * \pi(A) = 0 \,,
 }
where $K$, $\bar{K}$, $\pi$, $dz^2$, $d\bar{z}^2$, and $*$ are defined in the paragraphs below.  The phase $p$ is $1$ for the so-called type~A theory, dual to the $O(N)$ model \cite{Klebanov:2002ja} and $i$ for type~B, dual to the Gross-Neveu model \cite{Sezgin:2003pt}; correspondingly, $\bar{p}=1$ or $-i$.

The components of $A$, and also $B$, are functions of the usual four bosonic coordinates $x^\mu$ together with spinorial oscillator coordinates (also bosonic) $Y^A = (y^\alpha,\bar{y}^{\dot\alpha})$ and $Z^A = (z^\alpha,\bar{z}^{\dot\alpha})$, where $\alpha$ and $\dot\alpha$ are doublet indices for the irreducible spinor representations of $SO(3,1)$.  The coordinates $Y^A$ do not participate in the differential structure of the theory: in other words, the exterior derivative $d$ acts only on $x^\mu$ and $Z^A$, and we never encounter one-forms $dY^A$.  $A$ and $B$ admit series expansions in $Y^A$ and $Z^A$.  Roughly speaking, the metric and spin connection come from the terms in $A$ that are quadratic in the $Y^A$ coordinates, while the part of $B$ which depends only on the $x^\mu$ is identified as a scalar field.

To formulate the equations, one uses an associative star product, defined by
 \eqn{StarDef}{
  f(Y,Z) * g(Y,Z) = {\cal N} \int d^4 u \, d^4 v \,
    f(Y+U,Z+U) g(Y+V,Z-V) e^{U^A V_A} \,,
 }
where the normalization factor ${\cal N}$ is such that $f * 1 = f$.  Indices are raised and lowered according to
 \eqn{RaiseLower}{
  U^A = \Omega^{AB} U_B \qquad\qquad
  U_A = U^B \Omega_{BA} \,.
 }
Here
 \eqn{OmegaDef}{
  \Omega_{AB} = \Omega^{AB} = \begin{pmatrix} \epsilon_{\alpha\beta} & 0 \\
     0 & \epsilon_{\dot\alpha\dot\beta} \end{pmatrix}
 }
and
 \eqn{epsDef}{
  \epsilon_{\alpha\beta} = \epsilon^{\alpha\beta} = \epsilon_{\dot\alpha\dot\beta} = 
    \epsilon^{\dot\alpha\dot\beta} = \begin{pmatrix} 0 & 1 \\ -1 & 0 \end{pmatrix} \,.
 }
The star product is associative, and
 \eqn{YZstar}{\seqalign{\span\TL & \span\TR &\qquad\qquad \span\TL & \span\TR}{
  Y^A * Y^B &= Y^A Y^B + \Omega^{AB} &
  Z^A * Z^B &= Z^A Z^B - \Omega^{AB}  \cr
  Y^A * Z^B &= Y^A Z^B - \Omega^{AB} &
  Z^A * Y^B &= Z^A Y^B + \Omega^{AB} \,.
 }}
The Kleinians
 \eqn{Kleinians}{
  K \equiv e^{z^\alpha y_\alpha} \qquad\qquad
  \bar{K} \equiv e^{\bar{z}^{\dot\alpha} \bar{y}_{\dot\alpha}}
 }
satisfy $K * K = \bar{K} * \bar{K} = 1$, and also
 \eqn{Kstar}{
  f(y,\bar{y};z,\bar{z}) * K = K f(-z,\bar{y};-y,\bar{z}) \qquad
  K * f(y,\bar{y};z,\bar{z}) = K f(z,\bar{y};y,\bar{z}) \,.
 }
The map $\pi$, and a closely related map $\bar\pi$, are defined by
 \eqn{piMap}{
  \pi(f(y,\bar{y};z,\bar{z};dz,d\bar{z})) &= f(-y,\bar{y};-z,\bar{z};-dz,d\bar{z})  \cr
  \bar\pi(f(y,\bar{y};z,\bar{z};dz,d\bar{z})) &= f(y,-\bar{y};z,-\bar{z};dz,-d\bar{z}) \,.
 }
For zero-forms (i.e.~cases where $f$ doesn't depend on $dz$ or $d\bar{z}$), we have $\pi(f) = K * f * K$ as a consequence of \eno{Kstar}.  We also define
 \eqn{dzTwo}{
  dz^2 = {1 \over 2} dz^\alpha \wedge dz_\alpha = -dz^1 \wedge dz^2 \qquad\qquad
  d\bar{z}^2 = {1 \over 2} d\bar{z}^{\dot\alpha} \wedge d\bar{z}_{\dot\alpha} = 
    -d\bar{z}^{\dot{1}} \wedge d\bar{z}^{\dot{2}} \,.
 }
All definitions needed in \eno{eoms} are now explicit.

Passing locally to a gauge where the higher spin spacetime connection $w$ vanishes, the higher spin equations take the form
 \eqn[c]{WzeroEqns}{
  d_Z s + s * s = p (b * K) dz^2 + \bar{p} (b * \bar{K}) d\bar{z}^2  \cr
  d_Z b + s * b - b * \pi(s) = 0
 }
where $s = s_\alpha dz^\alpha + \bar{s}_{\dot\alpha} d\bar{z}^{\dot\alpha}$ is the spinorial part of the gauge field, and $b$, $s_\alpha$, and $\bar{s}_{\dot\alpha}$ are now functions only of $Y^A$ and $Z^A$.  Dependence on $x^\mu$ is prevented by the $x^\mu$ components of the full equations of motion \eno{eoms} in the $w=0$ gauge.  By $d_Z$ we mean the exterior derivative with respect to only the $Z^A$ variables; likewise, $d_x$ refers to the exterior derivative with respect to only the $x^\mu$ variables.  We use lowercase $b$ and $s$ in $w=0$ gauge so as to distinguish these quantities from their images in a more general gauge.

\section{The ansatz}
\label{ANSATZ}

In components, the equations \eno{WzeroEqns} read
 \eqn{WZEcomponents}{
  {\partial s_2 \over \partial z^1} - 
    {\partial s_1 \over \partial z^2} + [s_1,s_2]_* &= -p(b * K)  \cr
  {\partial b \over \partial z^\alpha} + s_\alpha * b + b * \pi(s_\alpha) &= 0  \cr
  {\partial \bar{s}_{\dot{2}} \over \partial \bar{z}^{\dot{1}}} - 
    {\partial \bar{s}_{\dot{1}} \over \partial \bar{z}^{\dot{2}}} + 
     [\bar{s}_{\dot{1}},\bar{s}_{\dot{2}}]_* &= -\bar{p}(b * \bar{K})  \cr
  {\partial b \over \partial \bar{z}^{\dot\alpha}} + \bar{s}_{\dot\alpha} * b -
    b * \pi(\bar{s}_{\dot\alpha}) &= 0  \cr
  {\partial \bar{s}_{\dot\beta} \over \partial z^\alpha} - 
    {\partial s_\alpha \over \partial \bar{z}^{\dot\beta}} + 
    [s_\alpha,\bar{s}_{\dot\beta}]_* &= 0 \,,
 }
where $[f,g]_* = f*g - g*f$.  Let's assume
 \eqn{TwoAnsatz}{
  s_1 = 0 = \bar{s}_{\dot{1}} \qquad\qquad 
  {\partial s_2 \over \partial \bar{z}^{\dot{2}}} = 0 = 
     {\partial \bar{s}_{\dot{2}} \over \partial z^2} \qquad\qquad
  {\partial b \over \partial Z^A} = 0 \,.
 }
These choices are convenient because the equations \eno{WZEcomponents} reduce to
 \eqn[c]{FiveEquations}{
  {\partial s_2 \over \partial z^1} = -p(b * K) \qquad\qquad
  {\partial \bar{s}_{\dot{2}} \over \partial \bar{z}^{\dot{1}}} = -\bar{p}(b * \bar{K})  \cr
  \{ s_2, b * K \}_* = 0 \qquad\qquad
  [\bar{s}_{\dot{2}},b * K]_* = 0 \qquad\qquad
  [s_2,\bar{s}_{\dot{2}}]_* = 0 \,,
 }
where $\{ f,g \}_* = f*g + g*f$.  Given $b = b(Y^A)$, we can immediately solve the first two equations in \eno{FiveEquations}:
 \eqn{sTwoSolve}{
  s_2 &= \int_0^1 dt \, \sigma_2(t) \qquad\hbox{where}\qquad
   \sigma_2(t) = -p z^1 \left[ b * K \right]_{z^1 \to t z^1}  \cr
  \bar{s}_{\dot{2}} &= \int_0^1 d\tilde{t} \, \bar\sigma_{\dot{2}}(\tilde{t}) 
    \qquad\hbox{where}\qquad
   \bar\sigma_{\dot{2}}(\tilde{t}) = 
     -\bar{p} \bar{z}^{\dot{1}} \left[ b * \bar{K} \right]_{\bar{z}^{\dot{1}} \to \tilde{t}
         \bar{z}^{\dot{1}}} \,.
 }
Note that the holomorphy conditions ${\partial s_2 \over \partial \bar{z}^{\dot{2}}} = 0 = {\partial \bar{s}_{\dot{2}} \over \partial z^2}$ which we assumed in \eno{TwoAnsatz} are automatically satisfied by \eno{sTwoSolve}.  Starting with $b = b(Y^A)$ and extracting $S$ through an integration similar to \eno{sTwoSolve} is a standard beginning to the perturbative approach of solving \eno{WzeroEqns}: See for example \cite{Sezgin:2005pv,Giombi:2010vg}.  The assumptions \eno{TwoAnsatz} make this perturbative approach exact.  However, the quadratic constraints in the second line of \eno{FiveEquations} must still be checked, and they do not hold for arbitrary functional forms $b(Y^A)$.  Before we indicate some functional forms $b(Y^A)$ for which the quadratic constraints {\it do} hold, let's note two final points.  First, by design, the forms \eno{sTwoSolve} are consistent with the requirement $S_A \to 0$ as $Z^A \to 0$, which is a standard gauge choice.  Second, we could generalize \eno{sTwoSolve} without spoiling the holomorphy conditions or this standard gauge choice by adding to $s_2$ a function only of $z^2$ and $Y^A$ which vanishes as $z^2 \to 0$; and likewise we could add to $\bar{s}_{\dot{2}}$ a function of $\bar{z}^{\dot{2}}$ and $Y^A$ which vanishes as $\bar{z}^{\dot{2}} \to 0$.  We will not consider such generalizations in this paper, but instead restrict ourselves to \eno{sTwoSolve} as written.

The simplest non-trivial solution to \eno{FiveEquations}-\eno{sTwoSolve} is
 \eqn{BOne}{
  b = b_0 \qquad\qquad
  \sigma_2(t) = -p b_0 z^1 e^{-tz^1 y^2 + z^2 y^1} \qquad\qquad
  \bar\sigma_{\dot{2}}(\tilde{t}) = -\bar{p} b_0 \bar{z}^{\dot{1}} 
    e^{-\tilde{t} \bar{z}^{\dot{1}} \bar{y}^{\dot{2}} + 
      \bar{z}^{\dot{2}} \bar{y}^{\dot{1}}} \,,
 }
where $b_0$ is a constant.  A stronger, unintegrated form of the quadratic constraints in \eno{FiveEquations} can be shown to hold for this case:
 \eqn{UnintegratedConstraints}{
  \{ \sigma_2(t), b * K \}_* = 0 \qquad\qquad
  [ \bar\sigma_{\dot{2}}(\tilde{t}), b * K ]_* = 0 \qquad\qquad
  [ \sigma_2(t), \bar\sigma_{\dot{2}}(\tilde{t}) ]_* = 0
 }
for all $t$ and $\tilde{t}$.  The second and third of these equations are trivially satisfied because $\sigma_2(t)$ and $b*K$ are fully holomorphic in $Y$ and $Z$, while $\bar\sigma_{\dot{2}}(\tilde{t})$ is fully anti-holomorphic.  The general result \eno{Kstar} implies in particular that $K$ anti-commutes with $y^\alpha$ and $z^\alpha$; so it is easy to see that it anti-commutes with $\sigma(t)$ as written in \eno{BOne}.  The case of constant $b$ case studied previously in \cite{Sezgin:2005pv}.  There however the authors imposed an $SO(3,1)$ symmetry, which lead to the constraint $s_\alpha = z_\alpha s(u)$ where $u=y^\alpha z_\alpha$ and $s(u)$ was expressed as an integral transform of confluent hypergeometric functions.  It is not clear to us that the solution of \cite{Sezgin:2005pv} is gauge-equivalent to ours.

An interesting generalization of the constant $b$ solution is
 \eqn{BTwo}{
  b = Q \, e^{q_{AB} Y^A Y^B} + R \, e^{r_{AB} Y^A Y^B}
 }
where the only non-vanishing components of $q_{AB}$ and $r_{AB}$ are those with $A$ and $B$ taking values in $\{1,\dot{1}\}$.  $Q$, $R$, and the non-zero components of $q_{AB}$ and $r_{AB}$ are parameters of the solution.  Straightforward but tedious computations suffice to show that the unintegrated constraints \eno{UnintegratedConstraints} are satisfied.  The importance of being able to take linear combinations of these special Gaussian solutions is that we need not stop at two terms: we can take arbitrarily many, or an integral of infinitely many.  In short, any function
 \eqn{BThree}{
  b = b((y^1)^2,y^1\bar{y}^{\dot{1}},(\bar{y}^{\dot{1}})^2)
 }
together with $s_2$ and $\bar{s}_{\dot{2}}$ as specified in \eno{sTwoSolve}, provides a solution of the equations \eno{WzeroEqns}.  A commonly imposed projection condition on field configurations restricts to functions $B$ which are invariant under sending $y \to iy$ and $\bar{y} \to -i\bar{y}$.  In the presence of this requirement, which is related to requiring only even integer spins in the full theory, $B$ must be a function of $(y^1)^4$, $y^1 \bar{y}^{\dot{1}}$, and $(\bar{y}^{\dot{1}})^4$.

Another interesting generalization of the constant $b$ solution is
 \eqn{BFour}{
  b = Q \, e^{q_{\alpha\dot\beta} y^\alpha \bar{y}^{\dot\beta}} \,,
 }
where $Q$ and the $q_{\alpha\dot\beta}$ are parameters.  As before, the unintegrated constraints \eno{UnintegratedConstraints} are satisfied once one imposes \eno{sTwoSolve}.  A caveat on solutions of the form \eno{BFour} is that if $\det q_{\alpha\dot\beta}$ is a real number less than or equal to $-1$ then some of the requisite star products are ill-defined, so the status of the solution is less clear.  There appears to be no general superposition principle for solutions of the form \eno{BFour} analogous to \eno{BTwo}.

\section{Gauge transformations and a mass deformation}
\label{GAUGE}

A trivial solution to Vasiliev's equations is $w=s=b=0$.  The $AdS_4$ solution, which we review in section~\ref{SPACETIME}, is gauge equivalent to this trivial solution.  We go on in section~\ref{SPACETIME} to explain in how to apply the same gauge transformation to other solutions starting in the $w=0$ gauge.  We then work out a particular example in section~\ref{EXAMPLE} in which $B \propto \zeta e^{y^1 \bar{y}^{\dot{1}} - y^2 \bar{y}^{\dot{2}}}$, where $\zeta$ is the radial coordinate in the Poincar\'e patch of $AdS_4$.  This example is interesting because the $B$ dependence just mentioned is, in the linearized theory, associated with a massive deformation of the $O(N)$ model.

\subsection{The spacetime connection}
\label{SPACETIME}

Let's review how the spacetime metric and spin connection are packaged into the spatial components $W$ of the higher spin gauge field $A$.  Starting from the vierbein $e^m = e^m_\mu dx^\mu$ and spin connection $\omega_{mn} = \omega_{\mu mn} dx^\mu$, we define
 \eqn{ew}{
  e_{\alpha\dot\beta} = {1 \over 2L} e^m \sigma_{m\alpha\dot\beta} \qquad\qquad
  \omega_{\alpha\beta} = {1 \over 2} \omega_{mn} \sigma^{mn}_{\alpha\beta} \qquad\qquad
  \bar\omega_{\dot\alpha\dot\beta} = -{1 \over 2} \omega_{mn} 
    \bar\sigma^{mn}_{\dot\alpha\dot\beta}
 }
and
 \eqn{ewTo}{
  e = {1 \over 2} e_{\alpha\dot\beta} y^\alpha \bar{y}^{\dot\beta} \qquad\qquad
  \omega = {1 \over 4} \omega_{\alpha\beta} y^\alpha y^\beta + 
   {1 \over 4} \bar\omega_{\dot\alpha\dot\beta} \bar{y}^{\dot\alpha} \bar{y}^{\dot\beta} \,.
 }
We have defined
 \eqn[c]{SigmaMatrices}{
  \sigma^m_{\alpha\dot\beta} = ({\bf 1},\vec\sigma) \qquad\qquad
  \bar\sigma^{m\dot\alpha\beta} = ({\bf 1},-\vec\sigma)  \cr
  \sigma^{mn}{}_\alpha{}^\beta = 
   {1 \over 4} (\sigma^m_{\alpha\dot\gamma} \bar\sigma^{n\dot\gamma\beta} -
     \sigma^n_{\alpha\dot\gamma} \bar\sigma^{m\dot\gamma\beta}) \qquad
  \bar\sigma^{mn\dot\alpha}{}_{\dot\beta} = 
   {1 \over 4} (\bar\sigma^{m\dot\alpha\gamma} \sigma^n_{\gamma\dot\beta} -
     \bar\sigma^{n\dot\alpha\gamma} \sigma^m_{\gamma\dot\beta}) \,,
 }
where $\vec\sigma$ are the usual Pauli matrices.  We express $AdS_4$ in Poincar\'e patch coordinates:
 \eqn{AdSFour}{
  e_{(0)}^m = \delta^m_\mu {L \over \zeta} dx^\mu
 }
with
 \eqn{AdSFourOmega}{
  \omega^{(0)}_{t\zeta} = {dt \over \zeta} \qquad\qquad
  \omega^{(0)}_{x^1\zeta} = -{dx^1 \over \zeta} \qquad\qquad
  \omega^{(0)}_{x^2\zeta} = -{dx^2 \over \zeta}
 }
and all other components of the spin connection vanishing except as required by the antisymmetry condition $\omega_{mn} = -\omega_{nm}$.  It is straightforward to check that
 \eqn{WzeroDef}{
  W_{(0)} = e_{(0)} + \omega_{(0)}
 }
satisfies the higher spin equations of motion with $S = B = 0$: That is, 
 \eqn{dWeq}{
  dW_{(0)} + W_{(0)} * W_{(0)} = 0 \,.
 }

In order to produce a more interesting solution of the equations of motion \eno{eoms}, we are going to to gauge transform one of our $w=0$ solutions.  Starting with a configuration $(a,b)$ of higher spin fields, the general gauge transformation to another configuration $(A,B)$ takes the form
 \eqn{GaugeOne}{
  d + A = g^{-1} * (d + a) * g \qquad\qquad
  B = g^{-1} * b * \pi(g) \,,
 }
where $g$ is a function of $x^\mu$, $Y^A$, and $Z^A$.  A more explicit form of the transformation of the gauge fields is 
 \eqn{GaugeTwo}{
  W = g^{-1} * d_x g + g^{-1} * w * g \qquad
  S = g^{-1} * d_Z g + g^{-1} * s * g \,.
 }
Our focus will be to set $w=0$.

The flatness of $W_{(0)}$ indicates that the $AdS_4$ solution is related to the trivial solution $w_{(0)} = 0$, $s_{(0)} = 0$, $b_{(0)} = 0$ by a gauge transformation.  For $(t,x^1,x^2) = (0,0,0)$, the gauge function may be represented as
 \eqn{gForm}{
  g^{\pm 1} = L^{\pm 1} \equiv {4 \over \sqrt{\zeta_0/\zeta} + 2 + \sqrt{\zeta/\zeta_0}} 
    \exp\left\{\mp {1 - \sqrt{\zeta/\zeta_0} \over 1 + \sqrt{\zeta/\zeta_0}} \, 
      \sigma^\zeta_{\alpha\dot\beta} y^\alpha \bar{y}^{\dot\beta} \right\} \,,
 }
where $\zeta_0$ is a parameter.  For a more complete description of this gauge transformation, including the full $x^\mu$ dependence, see for example \cite{Giombi:2010vg}.

\subsection{An example}
\label{EXAMPLE}

As an example of the procedure outlined in the previous section, let's consider the solution
 \eqn[c]{bEx}{
  b = b_0 e^{-\lambda (y^1 \bar{y}^{\dot{1}} - y^2 \bar{y}^{\dot{2}})}  \cr
  \sigma_2(t) = -p b_0 z^1 e^{(y^1-\lambda\bar{y}^{\dot{2}}) z^2 - 
    t (y^2-\lambda\bar{y}^{\dot{1}}) z^1} \qquad\qquad
  \bar\sigma_{\dot{2}}(\tilde{t}) = -\bar{p} b_0 \bar{z}^{\dot{1}} 
   e^{(\bar{y}^{\dot{1}}-\lambda y^2) \bar{z}^{\dot{2}} - 
    \tilde{t} (\bar{y}^{\dot{2}}-\lambda y^1) \bar{z}^{\dot{1}}} \,,
 }
where $b_0$ and $\lambda$ are real parameters.\footnote{The solution \eno{bEx} obeys the projection conditions that complete the characterization of the minimal higher spin theories, provided $b_0$ and $\lambda$ are real.  In the notation of \cite{Giombi:2012ms}, these projections are $\pi(\bar\pi(X)) = X$ for $X = W$, $S$, and $B$, together with $\iota_+(W) = -W$, $\iota_+(S) = -S$, and $\iota_-(B) = B$, where $\iota_\pm$ are linear maps which reverse the order of star products and send $(y,\bar{y},z,\bar{z},dz,d\bar{z}) \to (iy,\pm i\bar{y},-iz,\mp i\bar{z},-idz,\mp id\bar{z})$.}  In making the gauge transformation, we choose $\sigma^\zeta = \sigma^3 = \tiny \begin{pmatrix} 1 & 0 \\ 0 & -1 \end{pmatrix}$, and this choice is in some sense ``diagonal'' with respect to our earlier choice of $s_2$ and $\bar{s}_{\dot{2}}$ as the preferred components of the gauge field.  Nothing prevents us from making a different choice of $\sigma^\zeta$, but the resulting solution would then be more complicated.

The easiest field to pass through the gauge transformation is $B$, and one finds, at $(t,x^1,x^2)=(0,0,0)$, that
 \eqn{Bphys}{
  B = {4b_0 \zeta_0 \over \lambda_+^2\zeta}
   \, e^{-(y^1 \bar{y}^{\dot{1}} - y^2 \bar{y}^{\dot{2}}) {\lambda_- \over \lambda_+}} \,,
 }
where we have defined combinations
 \eqn{lambdapm}{
  \lambda_\pm = 1+\lambda \pm (1-\lambda) \zeta_0/\zeta
 }
which come up repeatedly after the gauge transformation.  We are interested in taking a $\zeta_0 \to \infty$ limit, because in this limit $B$ becomes translationally invariant in the boundary directions.  (Another way to put this is that boundary variation of $B$ takes place over a length scale $\Delta x \sim \zeta_0$, and we are taking that length scale to infinity.)  The specific limit we will consider is $\epsilon \to 0$ where
 \eqn{SpecialLimit}{
  \lambda = 1 - 2\epsilon \qquad\qquad
  \zeta_0 = {1 \over \epsilon^2}
 }
with $b_0$ held constant.  Passing \eno{Bphys} through this limit, we find
 \eqn{FinalB}{
  B = b_0 \zeta e^{y^1 \bar{y}^{\dot{1}} - y^2 \bar{y}^{\dot{2}}} \,.
 }
The scalar field in the higher spin theory is
 \eqn{FinalPhi}{
  \phi \equiv B\Big|_{Y^A=0} = b_0 \zeta \,.
 }

The spinor part of the gauge field may be expressed as
 \eqn{SaExpress}{
  S_2 = \int_0^1 du \, \Sigma_2(u)
 }
where
 \eqn{SigmaDef}{
  \Sigma_2(u) = {dt \over du} L^{-1} * \sigma_2(t) * L \,,
 }
and $u = u(t)$ is a conveniently chosen integration variable, with $u(0) = 0$ and $u(1) = 1$.  In the present case, a convenient definition is
 \eqn{uDef}{
  u = {t \lambda_+ \over 2(1-t) \sqrt{\zeta_0/\zeta} + t \lambda_+} \,,
 }
because then one finds 
 \eqn{SigmaExpress}{
  \Sigma_2(u) = -{4pb_0 \zeta_0/\zeta \over \lambda_+^2} z^1
    \exp\left\{ \left( y^1 - {\lambda_- \over \lambda_+} \bar{y}^{\dot{2}} \right)
      z^2 - u \left( y^2 - {\lambda_- \over \lambda_+} \bar{y}^{\dot{1}} \right) z^1
      \right\} \,.
 }
Similar expressions can be found for $\bar{S}_{\dot{2}} = \int_0^1 d\tilde{u} \, \bar\Sigma_{\dot{2}}(\tilde{u})$.  As before, these expressions are valid only at $(t,x^1,x^2) = (0,0,0)$; however, we may impose \eno{SpecialLimit} and pass to the $\epsilon \to 0$ limit to obtain the translationally invariant expressions
 \eqn{FinalSigmas}{
  \Sigma_2(u) = -p b_0 \zeta z^1 e^{-u (y^2 + \bar{y}^{\dot{1}}) z^1 + 
    (y^1 + \bar{y}^{\dot{2}}) z^2} \qquad\qquad
  \bar\Sigma_{\dot{2}}(\tilde{u}) = -\bar{p} b_0 \zeta \bar{z}^{\dot{1}} 
    e^{-\tilde{u} (\bar{y}^{\dot{2}} + y^1) \bar{z}^{\dot{1}} + 
      (\bar{y}^{\dot{1}} + y^2) \bar{z}^{\dot{2}}} \,.
 }

It is possible to check directly that the full equations of motion \eno{eoms} are satisfied when we set
 \eqn[c]{FullSolnForm}{
  B = b_0 \zeta e^{y^1 \bar{y}^{\dot{1}} - y^2 \bar{y}^{\dot{2}}}
    \qquad\qquad W = W_{(0)}  \cr
  S_1 = \bar{S}_{\dot{1}} = 0
    \qquad\qquad S_2 = \int_0^1 du \, \Sigma_2(u)
    \qquad\qquad \bar{S}_{\dot{2}} = \int_0^1 d\tilde{u} \, 
      \bar\Sigma_{\dot{2}}(\tilde{u})
 }
with $\Sigma_2(u)$ and $\bar\Sigma_{\dot{2}}(\tilde{u})$ as given in \eno{FinalSigmas}, and with the $AdS_4$ connection $W_{(0)}$ as defined in \eno{WzeroDef}.  However, there is an important subtlety: star products of $\Sigma_2(u)$ with $B$, which come up in the $D_{z^2} B = 0$ component of the equations of motion, formally diverge once one has passed to the translationally invariant limit; however, if one replaces $\Sigma_2(u)$ by $\Sigma_2(t,u) \equiv \Sigma_2(u) \big|_{z^2 \to t z^2}$, then $D_{z^2} B$ is proportional to $\{ \Sigma_2(t,u), B*K \}_*$, which vanishes identically.  A similar regulator is needed in order to check the equation $D_{\bar{z}^{\dot{2}}} B = 0$.  The other equations of motion can be handled without recourse to this type of regulator.  We caution that in other gauges, field configurations involving projectors such as $e^{y^1 \bar{y}^{\dot{1}} - y^2 \bar{y}^{\dot{2}}}$ often lead to divergences, for instance in $F_{z^1z^2}$, which do not cancel.  Thus it is challenging to find a solution analogous to \eno{FullSolnForm} in a covariant gauge.

The solution \eno{FullSolnForm} is interesting because in a linearization around $AdS_4$, the natural interpretation of the scalar profile \eno{FinalB} and \eno{FinalPhi} is that one is deforming the dual $O(N)$ field theory by a constant mass term for the $N$-dimensional vector $\vec\phi$: To see this, compare the scalar profile to the bulk-to-boundary propagators discussed, for example, in \cite{Giombi:2009wh,Vasiliev:2012vf,Didenko:2012tv}.  Once we introduce the spinorial connection based on \eno{FinalSigmas}, we obtain an exact generalization to the full non-linear equations of motion.  It is tempting to characterize this solution as a holographic dual of the massive $O(N)$ model.  However, caution is in order, because we do not fully understand how the explicit breaking of Lorentz symmetry inherent in our gauge choice $S_1=\bar{S}_{\dot{1}}=0$ affects the holographic interpretation.  Certainly it complicates the usual method \cite{Vasiliev:1999ba,Sezgin:2002ru} of extracting a privileged spacetime metric.\footnote{We thank S.~Didenko for a discussion on this point.}

\section{Conclusions}

The ansatz \eno{TwoAnsatz} in axial gauge significantly simplifies the equations of Vasiliev's higher spin theories in four dimensions, leading to a broad class of solutions for $b$ depending only on $y^1$ and $\bar{y}^{\dot{1}}$.  Privileging one component of a spinor over the other is in some settings related to picking out a null direction.  To see this, recall the equivalence of vectors and bi-spinors: $v_{\alpha\dot\beta} = v_m \sigma^m_{\alpha\dot\beta}$.  If we choose, for example, $v_m = (1,0,0,1)$, then $v_{\alpha\dot\alpha} y^\alpha \bar{y}^{\dot\alpha} = 2y^1 \bar{y}^{\dot{1}}$, showing that $y^1$ and $\bar{y}^{\dot{1}}$ have been privileged over $y^2$ and $\bar{y}^{\dot{2}}$.  Thus it is a reasonable guess that the solutions where $b = b((y^1)^2,y^1\bar{y}^{\dot{1}},(\bar{y}^{\dot{1}})^2)$ are related to shock waves, or to metrics expressed in terms of an Eddington-Finkelstein coordinate.  We hope to report further on this class of solutions in the future.

In a more limited but interesting class of solutions, $b$ depends on all four $Y^A$ variables, but only through the Gaussian expression given in \eno{BFour}.  We have explained how a simple special case, $b \propto e^{-\lambda (y^1 \bar{y}^{\dot{1}} - y^2 \bar{y}^{\dot{2}})}$, can be endowed with spacetime dependence through a gauge transformation.  In a suitable limit, this special case provides an exact solution improving upon the linearized description of a uniform mass deformation of the planar $O(N)$ model; note however that a cancellation of divergences is required in order to verify the $DB = 0$ equation.  It would clearly be of interest to compute two-point correlators in this higher spin geometry.  If indeed its interpretation as a dual of the massive $O(N)$ model is correct, then correlators should have a Lorentz invariant spectral weight with a continuum of states above a gap.  Additional solutions of the full Vasiliev equations \eno{eoms} might be constructed in a similar spirit; in particular, it is reasonable to suspect that an exact axial gauge solution might be available in which the spatial part of the connection $W$ is the same as for $AdS_4$, while the profile of the scalar master field $B$ is the $AdS_4$ bulk-to-boundary propagator.

Also important for future work is to generalize the Lorentz covariant treatment of the background metric to situations where as a matter of gauge choice one introduces parameters that break Lorentz symmetry.  Our gauge choice is of this type since it can be expressed as $\ell^\alpha S_\alpha = 0 = \bar\ell^{\dot\alpha} \bar{S}_{\dot\alpha}$ where $\ell^\alpha = {\tiny \begin{pmatrix} 1 \\ 0 \end{pmatrix}} = \bar\ell^{\dot\alpha}$, contrasting with the Lorentz-symmetric condition $z^\alpha S_\alpha = 0 = \bar{z}^{\dot\alpha} \bar{S}_{\dot\alpha}$ studied in previous works such as \cite{Vasiliev:1999ba,Sezgin:2002ru}.

\section*{Acknowledgments}

We are grateful to S.~Giombi for helpful discussions and especially to the referee for useful comments that led to a revision of section~\ref{GAUGE}.  This work was supported in part by the Department of Energy under Grant No.~DE-FG02-91ER40671.

\bibliographystyle{JHEP}
\bibliography{vasiliev}

\end{document}